\documentclass[11pt]{article}
\pdfoutput=1
\usepackage[centertags]{amsmath}
\usepackage[square, comma, sort&compress,numbers]{natbib}
\usepackage{array,multirow}
\numberwithin{equation}{section}
\usepackage{amssymb}
\usepackage{graphicx}
\usepackage{color}

\usepackage{epsfig}
\usepackage{bbold}

\usepackage{wrapfig}
\usepackage{float}
\usepackage{soul}

\usepackage{tkz-euclide}

\usepackage{tikz,pgf}
\usetikzlibrary{shapes}
\usetikzlibrary{calc}
\usetikzlibrary{decorations.pathmorphing}
\usetikzlibrary{decorations.pathreplacing,shapes.misc}
\usetikzlibrary{positioning}
\usetikzlibrary{arrows}
\usetikzlibrary{decorations.markings}
\usetikzlibrary{shadings}

\usetikzlibrary{intersections}

\def\be{\begin{equation}}
\def\ee{\end{equation}}
\def\ba{\begin{array}}
\def\ea{\end{array}}

\def\dps{\displaystyle}
\newcommand{\half}{\frac{1}{2}}

\def\1{\tilde{1}}
\def\2{\tilde{2}}
\def\3{\tilde{3}}

\newdimen\tableauside\tableauside=1.0ex
\newdimen\tableaurule\tableaurule=0.4pt
\newdimen\tableaustep
\def\phantomhrule#1{\hbox{\vbox to0pt{\hrule height\tableaurule
width#1\vss}}}
\def\phantomvrule#1{\vbox{\hbox to0pt{\vrule width\tableaurule
height#1\hss}}}
\def\sqr{\vbox{%
  \phantomhrule\tableaustep

\hbox{\phantomvrule\tableaustep\kern\tableaustep\phantomvrule\tableaustep}%
  \hbox{\vbox{\phantomhrule\tableauside}\kern-\tableaurule}}}
\def\squares#1{\hbox{\count0=#1\noindent\loop\sqr
  \advance\count0 by-1 \ifnum\count0>0\repeat}}
\def\tableau#1{\vcenter{\offinterlineskip
  \tableaustep=\tableauside\advance\tableaustep by-\tableaurule
  \kern\normallineskip\hbox
    {\kern\normallineskip\vbox
      {\gettableau#1 0 }%
     \kern\normallineskip\kern\tableaurule}%
  \kern\normallineskip\kern\tableaurule}}
\def\gettableau#1 {\ifnum#1=0\let\next=\null\else
  \squares{#1}\let\next=\gettableau\fi\next}

\newcommand{\bref}[1]{\textbf{\ref{#1}}}

\def\cF{\mathcal{F}}

\def\cO{\mathcal{O}}

\def\cV{\mathcal{V}}

\numberwithin{equation}{section} \makeatletter
\@addtoreset{equation}{section}

\def\be{\begin{equation}}
\def\ee{\end{equation}}
\def\ba{\begin{array}}
\def\ea{\end{array}}

\def\dps{\displaystyle}

\def\ba{\begin{array}}
\def\ea{\end{array}}

\def\dps{\displaystyle}

\usepackage{jheppub}
\makeatletter
\def\@fpheader{\vspace{-.1cm}}
\makeatother

\title{Example of $4$-pt non-vacuum $\mathcal{W}_3$ classical block}
\author[a]{Mikhail\ Pavlov}

\affiliation[a]{I.E. Tamm Department of Theoretical Physics, \\P.N. Lebedev Physical
Institute,\\ Leninsky ave. 53, 119991 Moscow, Russia}
\emailAdd{pavlov@lpi.ru}

\abstract{ In this note, we study a special case of the $4$-pt  non-vacuum classical block associated with the $\mathcal{W}_3$ algebra. We formulate the monodromy problem for the block and derive monodromy equations within the heavy-light approximation. Fixing the remaining functional arbitrariness using parameters of the  $4$-pt vacuum $\mathcal{W}_3$ block, we compute the $4$-pt  non-vacuum $\mathcal{W}_3$ block function.}

\makeatletter
\def\@fpheader{\vspace{-.1cm}}
\makeatother

\begin{document}

\maketitle
\flushbottom

\section{Introduction}

~~~~ Two-dimensional conformal theories with extended symmetry generated by spin-$N$, $N\geq3$  currents are a natural generalization of theories with the Virasoro algebra corresponding to $N=2$. The $\mathcal{W}_3$ symmetry was proposed for the minimal model $\mathcal{M}(6,5)$ \cite{Zamolodchikov:1985wn, Fateev:1987vh} which describes the three-state Potts model, and was extended to the theories with $N > 3$ \cite{Fateev:1987zh, Bouwknegt:1994zux}.\footnote{ $\mathcal{W}_N$ algebras and their supersymmetric generalizations (see \cite{Bergshoeff:1991dz}, for instance)  also arise when considering integrable models \cite{gel1978family}. } Moreover, the $W_{1+\infty}$ algebras appearing in the context of the higher spin theories may be considered as limiting cases of $W_N$ algebras \cite{10.1143/PTPS.118.343}. 

From the CFT$_2$ standpoint, the  main object of study in theories with $\mathcal{W}_N$ symmetry are the basis elements in the space of correlation functions - conformal blocks. In this paper, we focus on the $\mathcal{W}_3$ algebra and consider the non-vacuum $4$-pt $\mathcal{W}_3$ classical block within the monodromy method. The monodromy properties of the given block are governed by the fusion rules of the $\mathcal{W}_3$ algebra. In contrast to the Virasoro blocks, generally $\mathcal{W}_3$ blocks contain functional arbitrariness. In other words, it is necessary to fix this arbitrariness, which imposes additional requirements on a particular $\mathcal{W}_3$ CFT$_2$. However, the arbitrariness is absent for the $4$-pt $\mathcal{W}_3$ vacuum block \cite{deBoer:2014sna}. Considering the non-vacuum block as a certain deformation of the vacuum block, we discuss ways of fixing this arbitrariness in Section \bref{sec:4NV}.

The paper is organized as follows. In Section \bref{sec:FR} we discuss $\mathcal{W}_3$ algebra in the classical limit and study its degenerate operators along with associated fusion rules. Section \bref{sec:4NV} is devoted to the monodromy method for classical $\mathcal{W}_3$ blocks. Here, we focus on calculations of the special $4$-pt non-vacuum block using fusion rules derived in Section \bref{sec:FR}. Section \bref{sec:conclusion} summarizes our results. Appendix \bref{sec:V13} contains details about the monodromy method related with the degenerate operator $V_{(3,1)}$ of the Virasoro algebra.

\section{Fusion rules of the $\mathcal{W}_3$ algebra}
\label{sec:FR}

In this section, we consider CFT$_2$ with $\mathcal{W}_3$ symmetry and derive fusion rules associated with degenerate operators of the $\mathcal{W}_3$ algebra. 
\paragraph{$\mathcal{W}_3$ algebra and highest-weight representations.} 

 $\mathcal{W}_3$ symmetry is generated by two currents $T(z)$ and  $W(z)$, the first of which is the energy-momentum tensor, and the second corresponds to the spin-3 current. These currents can be decomposed as follows
\be
T(z) = \sum_{m=-\infty}^{\infty}\frac{L_m}{z^{m+2}} \;, \qquad W(z) = \sum_{n=-\infty}^{\infty}\frac{W_n}{z^{n+3}}\;,
\ee
where modes $L_{m}$ and $W_{n}$ satisfy commutation relations
\be
\ba{c}\label{AW3}
    \left[L_n,L_m\right]=(n-m)L_{n+m}+\dps \frac{c}{12}(n^3-n)
    \delta_{n+m,0}\;,
    \\
    \\
    \left[L_n,W_m\right]=(2n-m)W_{n+m}\;,
    \\
    \\
   \dps\left[W_n,W_m\right]=\frac{c}{3\cdot5!}(n^2-1)(n^2-4)n
    \delta_{n+m,0}+\frac{16}{22+5c}(n-m)\Lambda_{n+m}+\\
    \\
    \dps\frac{(n-m)}{30}\left( 2m^2 +2n^2 - m n -8  \right)L_{n+m}\;,
\ea
\ee
where 
\be
\label{nl}
\Lambda_{m} = \sum_{p \leq - 2} L_{p} L_{m-p} + \sum_{p \geq - 1} L_{m-p} L_{p} - \frac{3(m+2)(m+3)}{10} L_{m}\;,
\ee
and for a central element $c$ it is convenient to use the parametrization
\be
\label{li}
c = 2 \left(1 + 12 (b^{1}+b^{-1})^2\right).  
\ee 
The associative algebra given by the relations \eqref{AW3} is called the $\mathcal{W}_3$ algebra \cite{Zamolodchikov:1985wn,Fateev:1987vh}. Due to the presence of the nonlinear term  \eqref{nl} the $\mathcal{W}_3$ algebra is not a Lie algebra, but its generators satisfy the Jacobi identities \cite{Zamolodchikov:1985wn}. 

In this paper, we focus on the classical limit $c \rightarrow \infty$ (or $b \rightarrow 0$). In this limit, one can neglect the non-linear term \eqref{nl} in the commutators \eqref{AW3}. In what follows, we discuss \eqref{AW3} in the classical limit only.  Note that the resulting algebra has the global subalgebra $sl(3)$ spanned by $8$ generators $(L_{0}, L_{\pm 1}, W_{0}, W_{\pm 1}, W_{\pm 2})$. In addition, the $\mathcal{W}_3$ algebra is finitely generated, more precisely, the set $(L_{0}, L_{\pm 1}, L_{\pm2}, W_{0})$ forms the entire algebra through relations \eqref{AW3}.

The representation theory of the $\mathcal{W}_3$ algebra can be constructed in analogy with that of the Virasoro algebra. A $\mathcal{W}_3$ highest-weight vector  $| \Delta, Q \rangle$  satisfies the following conditions
\be
\label{HWW}
L_{n} | \Delta, Q \rangle = \delta_{n,0} ~ \Delta | \Delta, Q \rangle\;, \qquad W_{n} | \Delta, Q \rangle = \delta_{n,0} ~ Q | \Delta, Q \rangle\;, \qquad  n \geq 0\;,
\ee
where $\Delta$ stands for the conformal dimension, $Q$ denotes the spin-$3$ charge. The $\mathcal{W}_3$ module is spanned by the basis states\footnote{For more detailed analysis, see \cite{Belavin:2016qaa}, Appendix E. }
\be
\label{FB}
\ba{c}
\mathcal{L}_{-I}  | \Delta, Q \rangle  \equiv L_{-i_1}...L_{-i_k} W_{-j_1}... W_{-j_p}  | \Delta, Q \rangle, \qquad I \equiv \{ i_1, ...., i_k; j_1, ... , j_p\}, 
\\
\\
1 \leq i_1 \leq ...\leq i_{k}, \qquad 1 \leq j_1 \leq ...\leq j_{p}\;.
\ea
\ee

For states \eqref{HWW}, we can define the primary operator $\cO_{\Delta, Q}$ using \footnote{ The local symmetry of the  $\mathcal{W}_3$ CFT$_2$ is governed by two copies of the $\mathcal{W}_3$ algebra. In this section we focus only on the chiral part of the symmetry, which acts on operators in the variable $z$ and omit the $\bar z$ dependence. }
\be
\label{OSC}
| \Delta, Q \rangle = \lim_{z\rightarrow 0}\cO_{\Delta, Q}(z)|0\rangle. 
\ee
In contrast to the Virasoro algebra generators $L_m$, the modes $W_n$ cannot be expressed in terms of differential operators with respect to the variable $z$. The differential realization of \eqref{AW3} involves two additional operators $\hat{W}_{-1} \cO_{_{\Delta, Q}}(z)$ and $\hat{W}_{-2} \cO_{_ {\Delta, Q}}(z)$ \cite{Watts:1994zq}
\be
\label{rea}
\ba{c}
\mathcal{L}_{m} \cO_{_{\Delta, Q}}(z) = \left(z^{m+1} \partial_{z} + \Delta(m+1) z^{m}\right) \cO_{_{\Delta, Q}}(z), 
\\
\\
\mathcal{W}_{n} \cO_{_{\Delta, Q}}(z) = \left(\dps \frac{Q}{2} (m+2)(m+1) z^{m} + (m+2) z^{m+1} \hat{W}_{-1} + z^{m+2} \hat{W}_{-2}\right)\cO_{_{\Delta, Q}}(z).  
\ea
\ee
In what follows, we use a simplified form of the subscript for the primary operators $\cO_{1}$, where the index $1$ denotes $(\Delta_1, Q_1)$.
\paragraph{$2,3$-pt correlators.}
The $\mathcal{W}_3$ symmetry imposes certain restrictions on $n$-pt correlation functions of primary operators governed by conformal Ward identities. The ones with currents $T(z)$ and $W(z)$ have the form
\begin{align}\label{Lw}
     &\langle T(z)\cO_1(z_1)
     \dots \cO_n(z_n)\rangle=\sum_{i=1}^n
     \left(\frac{\Delta_i}{(z-z_i)^2}+\frac{\partial_i}{(z-z_i)}
     \right)
     \langle \cO_1(z_1)\dots \cO_n(z_n)\rangle,\\\label{Ww}
     &\langle W(z)\cO_1(z_1)\dots \cO_n(z_n)\rangle=\sum_{i=1}^n
     \left(\frac{Q_i}{(z-z_i)^3}+\frac{\hat{W}_{-1}^{(i)}}{(z-z_i)^2}+
     \frac{\hat{W}_{-2}^{(i)}}{(z-z_i)}\right)
     \langle \cO_1(z_1)\dots \cO_n(z_n)\rangle, 
\end{align} where we use the notation
\be
\hat{W}_{-p}^{(i)} \langle \cO_1(z_1)\dots \cO_n(z_n)\rangle \equiv \langle \cO_1(z_1)\dots  \hat{W}_{-p} \cO_i(z_i) \dots \cO_n(z_n)\rangle, \qquad p = 1,2.  
\ee
The asymptotic behavior of the currents at $z \rightarrow \infty$ is given by
\be
\label{asym}
T(z) \rightarrow z^{-4} \;, \qquad W(z) \rightarrow z^{-6} \;, \quad \text{at} \quad z\rightarrow \infty.
\ee
Applying \eqref{asym} to \eqref{Lw} and \eqref{Ww}, one obtains $8$ global Ward identities. Three of them associated with $T(z)$ have the form
\be
\label{stY}
\ba{c}
\dps \sum^{n}_{i=1} \partial_{i} \langle \cO_{1}(z_1) \dots \cO_n(z_n) \rangle  =0, \\
\dps \sum^{n}_{i=1} (z_i \partial_{i} + \Delta_i) \langle \cO_{1}(z_1) \dots\cO_n(z_n) \rangle  =0, \\
\dps \sum^{n}_{i=1} (z^2_i \partial_{i} + 2 z_i \Delta_i) \langle \cO_{1}(z_1) \dots \cO_n(z_n) \rangle  =0.
\ea
\ee
Five remaining identities following from \eqref{Ww} are 
\be
\label{extY}
\ba{c}
\dps \sum^{n}_{i=1} \hat{W}_{_{-2}}^{(i)} \langle \cO_{1}(z_1) \dots \cO_n(z_n) \rangle  =0, \\
 \dps \sum^{n}_{i=1} \left(z_i \hat{W}_{_{-2}}^{(i)} +  \hat{W}^{(i)}_{_{-1}} \right) \langle \cO_{1}(z_1) \dots \cO_n(z_n) \rangle =0, \\
\dps \sum^{n}_{i=1} \left(z^2_i \hat{W}_{_{-2}}^{(i)} + 2 z_i \hat{W}^{(i)}_{_{-1}} + Q_{i}\right) \langle \cO_{1}(z_1) \dots \cO_n(z_n) \rangle =0, \\
 \dps \sum^{n}_{i=1} \left(z^3_i \hat{W}_{_{-2}}^{(i)} + 3 z^2_i \hat{W}^{(i)}_{_{-1}} + 3 z_i Q_{i}\right) \langle \cO_{1}(z_1) \dots \cO_n(z_n) \rangle =0, \\
  \dps \sum^{n}_{i=1} \left(z^4_i \hat{W}_{_{-2}}^{(i)} + 4 z^3_i \hat{W}^{(i)}_{_{-1}} + 6 z^2_i Q_{i}\right) \langle \cO_{1}(z_1) \dots \cO_n(z_n) \rangle =0. 
\ea
\ee

The 2-pt correlation function is fixed (up to normalization) by the Ward identities \cite{Fateev:2007ab}
\be
\label{2pt}
\langle \cO_{1}(z_1) \cO_{2}(z_2) \rangle = \frac{\delta_{\Delta_1, \Delta_2} \delta_{Q_1, -Q_2}}{(z_1-z_2)^{2\Delta_1}}\;,
\ee
so the $\mathcal{W}_3$ symmetry imposes $\Delta_1 = \Delta_2$ and $Q_1 + Q_2=0$. In contrast to the Virasoro algebra, conditions \eqref{stY} and \eqref{extY} do not completely determine the $3$-pt correlation function. Indeed, the $3$-pt function depends on nine variables: coordinates $z_j$ and  $\hat{W}^{(j)}_{-1}, \hat{W}^{(j)}_{-2}$, $ j = 1, 2, 3$. So 8 Ward identities are not enough to completely eliminate functional arbitrariness.\footnote{Nevertheless, these functions are explicitly known in some special cases, see \cite{Fateev:2007ab, Fateev:2008bm, Fateev:2011qa}. } However, the conditions \eqref{stY} fix the coordinate dependence of the $3$-pt function, which has the form 
\be
\label{3pt}
V_{123} \equiv 
    \langle \cO_1(z_1) \cO_2 (z_2) 
    \cO_3 (z_3)\rangle = C_{123}
    (z_1-z_2)^{(\Delta_3-\Delta_3-\Delta_1)}(z_1-z_3)^{(\Delta_2-
    \Delta_3-\Delta_1)}
    (z_2-z_3)^{(\Delta_2+\Delta_3-\Delta_1)}
\ee
where $C_{123}$ is a structure constant.\footnote{Notice that the structure constant $C_{123}$ in \eqref{3pt} depends on $\hat{W}^{(j)}_{-1}, \hat{W}^{(j)}_{-2}, \; i=1,2,3$ and still is a subject of equations \eqref{extY}. Since the $3$-pt function contains functional arbitrariness, we do not resolve equations \eqref{extY} for $n=3$ explicitly and leave $3$-pt function to have the schematic form \eqref{3pt}.}

\paragraph{Null-vectors of the $\mathcal{W}_3$ algebra.}  Let $|\chi \rangle$ be a  highest-weight vector and $|\tilde \chi \rangle$ is some descendant state \eqref{FB} obtained from  $|\chi \rangle$. $|\tilde \chi \rangle$ is called a null-vector if it satisfies the following conditions  
\be
\label{sing}
L_{1}| \tilde{\chi} \rangle = L_{2}|\tilde{\chi} \rangle = W_{1}| \tilde{\chi} \rangle = 0\;,  
\ee This implies that the $\mathcal{W}_3$ module $\cV_{\chi}$  does not realize an irreducible representation of the $\mathcal{W}_3$ algebra.
But we can obtain one by factorizing over the null-vectors, i.e. $|\tilde \chi \rangle =0$.  The operator $\chi$ corresponding to $|\chi \rangle$ in a sense of \eqref{OSC} will be called degenerate.

For the further analysis in Section \bref{sec:4NV} we consider the vector $|\psi \rangle$  with the following conformal dimension and the spin-$3$ charge 
\be
\label{resc}
\dps \Delta_{\psi} = -1 - \frac{4 b^2}{3}, \qquad Q^2_{\psi} = - \frac{2 \Delta^2_{\psi}}{27} \frac{5 b + 3 b^{-1}}{3 b + 5 b^{-1}}.
\ee
 In this case, there are null-vectors at levels one, two and three which are given by \cite{Bajnok:1992nj, Fateev:1987vh, Watts:1994zq}
\be 
    |\psi_1 \rangle =  \left(W_{-1}-\frac{3Q_{\psi}}{2\Delta_{\psi}}L_{-1}\right)|\psi \rangle, 
    \ee
    \be
    |\psi_2 \rangle = \left(W_{-2}-\frac{6Q_{\psi}}{\Delta_{\psi}(5\Delta_{\psi}+1)}\left( L_{-1}^2 - (\Delta_{\psi}+1)L_{-2}\right) \right)|\psi \rangle,
\ee
\be\label{3null}
 |\psi_3 \rangle = \left(W_{-3}+ \frac{Q_{\psi}}{\Delta_{\psi}(5\Delta_{\psi}+1)} \left( -
  \frac{16}{\Delta_{\psi}+1}
    L_{-1}^3+ 12 L_{-1}L_{-2}+
    \frac{3(\Delta_{\psi}-3)}{2} L_{-3}\right)\right) 
    |\psi \rangle,
\ee

\vspace{2mm}

\noindent and due to factorization we set $ |\psi_3 \rangle =  |\psi_2 \rangle =  |\psi_1 \rangle = 0$. Hence, the correlation functions with operators  that correspond to null-vectors are equal to 0. It imposes conditions (fusion rules) on the conformal dimensions and  spin-$3$ charges of the remaining operators that appear in these correlation functions.

 Consider the $3$-pt correlation function $V_{\alpha \beta \psi_{_{3}}}$ where the operator $\psi_3$ is associated with the null-vector \eqref{3null}. Putting $V_{\alpha \beta \psi_{_{3}}}=0$ due to $| \psi_3 \rangle =0$ one finds \cite{Fateev:2007ab} 
\be \label{fg}
\ba{c} 
  32 Q_{\psi} (\Delta_{\alpha} - \Delta_{
  \beta})^3 - 12 Q_{\psi} (\Delta_{\psi}+1) (\Delta^2_{\alpha}-\Delta^2_{\beta}) - Q_{\psi} (15 \Delta^2_{\psi} - 12 \Delta_{\psi} -1 )(\Delta_{\alpha} - \Delta_{\beta}) + \\
  \\
  (\Delta^2_{\psi} + \Delta_{\psi})(5 \Delta_{\psi}+1) (Q_{\alpha} - Q_{\beta}) = 0 \;.
\ea
\ee
The equation \eqref{fg} defines the fusion rule connecting dimensions $\Delta_{\alpha, \beta}$ and the charges $Q_{\alpha, \beta}$ of the operators $\cO_{\alpha}$ and $\cO_{\beta}$. We concentrate on the classical limit, when all $\Delta$ and $Q$ are linear in the central charge at $c\rightarrow \infty$, which can also be rewritten using \eqref{li}
\be
\label{class}
\Delta = \frac{4 \epsilon}{b^2}, \qquad Q =  \frac{4 q}{b^2}, \qquad b \rightarrow 0, 
\ee
where $\epsilon, q$ are called classical dimensions/charges respectively. In order to analyze monodromy properties of the correlation functions with an insertion of the operator $\psi_{3}$ we  solve the equation \eqref{fg} in terms of a variable $\gamma = \Delta_{\beta} - \Delta_{\alpha} - \Delta_{\psi}$ at $b \rightarrow 0$. The solutions are
\be
\label{gs}
\ba{c}
\dps \gamma_{1} = 1 - \frac{R^{1/3}}{3^{2/3} Q_0} - \frac{Q_0(1-4\epsilon_{\alpha})}{3^{1/3} R^{1/3}}\;,
\\
\\
\dps \gamma_{2,3} = 1 + \left(-\frac{1}{3}\right)^{1/3} \frac{Q_0 (1-4\epsilon_{\alpha})}{R^{1/3}} + \left(\frac{1}{3}\right)^{1/3}\frac{(1\pm i \sqrt{3})R^{1/3}}{6 Q_0}\;, 
\ea
\ee
where
\be
R = \sqrt{3} \sqrt{ 3 Q_0^4 \left(  q_{\alpha}-q_{\beta}\right)^2 -Q_0^6 (1 - 4 \epsilon_{\alpha})^3} + 3 Q^2_{0} (q_{\alpha} - q_{\beta})^2\;,
\ee
and $Q^2_{0} = -2/45$.

The simplest fusion rule corresponds to the case $q_{\alpha} = q_{\beta}$ and reads
\be
\label{ngs}
\ba{c}
\gamma_1 = 1 \;, \qquad \gamma_{2,3} = \dps  \left(1 \pm  \sqrt{1 - 4 \epsilon_{\alpha}} \right)\;. 
\ea
\ee
Note that the conditions above do not depend on classical charges. In fact, fusion rules \eqref{ngs} exactly reproduce ones corresponding to a singular operator $V_{(3,1)}$ of the Virasoro algebra (see Appendix \bref{sec:V13} for details). Hence, the monodromy properties of the $3$-pt function are determined by the Virasoro algebra despite the fact that $q_{\alpha} \neq 0$. 

\section{$4$-pt $\mathcal{W}_3$ blocks: monodromy method} 
\label{sec:4NV}
\subsection{Classical conformal blocks and
the monodromy problem}
\label{sec:cl}

Here we consider $4$-pt $\mathcal{W}_3$ blocks in the classical limit within the monodromy method.

\paragraph{Classical conformal blocks.}
 
In 2d CFT with the $\mathcal{W}_3 \oplus \overline{\mathcal{W}}_3$ symmetry the $4$-pt correlation function of primary operators $\cO_{i} (z_i, \bar z_i), ~ i = \overline{1,4}$ can be decomposed into conformal blocks 
\be
\ba{c}
\label{deco}
\dps \langle \cO_1(z_1, \bar z_1) \cO_2 (z_2, \bar z_2) 
    \cO_3 (z_3, \bar z_3) \cO_4 (z_4, \bar z_4)\rangle = 
   \dps \sum_{p \in S}  C_{12p}  C_{p34} \Big| \cF_{W}\left(z_i| \Delta_i, Q_i,  \tilde{\Delta}_p, \tilde Q_p, c\right)\Big|^2\;, ~~~~~~~~~~~~\\
   \\
   \cF_{W}\left(z_i| \Delta_i, Q_i,  \tilde{\Delta}_p, \tilde Q_p, c\right) = \dps \frac{ \langle \cO_1(z_1, \bar z_1) \cO_2 (z_2, \bar z_2) (\mathbb{P}_{\tilde \Delta_p, \tilde Q_p} \otimes | \bar{\tilde \Delta}_p, \bar{\tilde Q}_p \rangle \langle \bar{\tilde \Delta}_p, \bar{\tilde Q}_p| ) \cO_3 (z_3, \bar z_3) \cO_4 (z_4, \bar z_4) \rangle}{   C_{12p}  C_{p34} \bar V(\bar z_1, \bar z_2, \bar z_3, \bar z_4)}, \\
   \\
   C_{abc}  = \langle \Delta_a, \bar \Delta_a, Q_a, \bar Q_a | \cO_b (1,1)| \Delta_c, \bar \Delta_c, Q_c, \bar Q_c \rangle \;, \quad \dps \mathbb{P}_{\tilde \Delta_p, \tilde Q_p} = \sum_{I} \frac{\mathcal{L}_{-I}| \tilde{\Delta}_p, \tilde Q_p \rangle \langle \tilde{\Delta}_p, \tilde Q_p | \mathcal{L}_{I}}{\langle \tilde{\Delta}_p, \tilde Q_p | \mathcal{L}_{I} \mathcal{L}_{-I}| \tilde{\Delta}_p, \tilde Q_p \rangle}\;,
 \ea
 \ee 
 where 
 \be
 \bar V(\bar z_1, \bar z_2, \bar z_3, \bar z_4) = \bar z^{\bar{\tilde \Delta}_p - \bar \Delta_1 - \bar \Delta_2}_{12} \bar z^{ \bar \Delta_1 -  \bar \Delta_2 - \bar{\tilde \Delta}_p}_2 \bar{z}^{ \bar \Delta_2 -  \bar \Delta_1 - \bar{\tilde \Delta}_p}_1 \bar z^{\bar{\tilde \Delta}_p - \bar \Delta_3 - \bar \Delta_4}_{34}, \qquad \bar{z}_{ij} \equiv \bar z_i - \bar z_j. 
 \ee
In \eqref{deco},  $\cF_{W}\left(z_i| \Delta_i, Q_i,  \tilde{\Delta}_p, \tilde Q_p, c\right)$ denotes a $4$-pt $\mathcal{W}_3$ conformal block, $\mathbb{P}_{\tilde \Delta_p, \tilde Q_p}$ is a projector on the $\mathcal{W}_3$ Verma module $\cV_{\tilde{\Delta}_p, \tilde Q_p}$ and $\mathcal{L}_{I}$ is given by \eqref{FB}. In the last line of \eqref{deco} $C_{abc}$ stands for structure constants determined using vectors $| \Delta, \bar \Delta, Q, \bar Q \rangle = | \Delta, Q\rangle \otimes | \bar \Delta, \bar Q \rangle$. The block singles out the contribution (more precisely, its holomorphic part) of the conformal family of the operator $\tilde{\cO}_p$.  We also insert the operator $| \bar{\tilde \Delta}_p, \bar{\tilde Q}_p \rangle \langle \bar{\tilde \Delta}_p, \bar{\tilde Q}_p |$ and introduce the function $\bar V$ to remove the residual antiholomorphic dependence and the contribution of structure constants from the block.  

 Unlike the Virasoro blocks, the  $\mathcal{W}_3$ conformal blocks are not completely determined by the symmetry algebra and conformal dimensions/spin-3 charges of external and internal operators. More precisely, it was shown \cite{B92, Wyllard:2009hg} that ratios
\be
\label{farb}
C_{n} = \frac{\langle \Delta_1, Q_1| \cO_{2}(1) (W_{-1})^n| \tilde{\Delta}, \tilde Q \rangle}{\langle \Delta_1, Q_1| \cO_{2}(1)| \tilde{\Delta}, \tilde Q \rangle}\;, \qquad \tilde{C}_{n} = \frac{\langle \tilde{\Delta}, \tilde Q| (W_{1})^n \cO_{3}(1) | \Delta_4,  Q_4 \rangle}{\langle \tilde{\Delta}, \tilde Q| \cO_{3}(1) | \Delta_4,  Q_4 \rangle}\;,
\ee
where $n = 1, 2, ...$ act as additional parameters for the conformal block.  In other words, the conformal block depends on two arbitrary functions, which correspond to two infinite sets of parameters \eqref{farb}. These two functions govern the functional arbitrariness of the conformal block. Note that in special cases this arbitrariness is fixed,\footnote{ The functional arbitrariness is also absent if we choose two external operators to be semi-degenerate, cf. \cite{Wyllard:2009hg, Fateev:2011qa}.} for example, for a vacuum block ($\tilde{\Delta} = \tilde{Q} = 0$), all parameters \eqref{farb} are equal to 0.

In the classical limit  $c\rightarrow \infty$,  assuming all conformal dimensions and spin-$3$ charges are proportional to $c$, it has been argued that the block exponentiates\footnote{In contrast to the case of Virasoro, where this property of the block was verified using recursion relations up to a sufficiently high order \cite{Zamolodchikov:1984eqp, Zamolodchikovv}  and was proved in \cite{Besken:2019jyw}, for the general $\mathcal{W}_3$ block it has the status of a plausible hypothesis (see also \cite{deBoer:2014sna, Hegde:2015dqh} for discussion in the context of AdS/CFT correspondence).}
\be
\lim_{c \to \infty} \cF_{W} \left(z_i| \Delta_i, Q_i,  \tilde{\Delta}, \tilde Q, c\right) = \exp\left[\frac{c}{6} f_{W} (z_i| \epsilon_i, \tilde \epsilon, q_i, \tilde q )\right]\;,
\ee
where $f_{W}(z_i| \epsilon, \tilde \epsilon, q, \tilde q)$ stands for the classical  block.  Classical dimensions/charges \eqref{class} are assumed to be finite in the limit $c \rightarrow \infty$. In what follows, we will omit the dependence on these parameters in the classical block function.

The monodromy method for calculating the $\mathcal{W}_3$ block is formulated by analogy to that for Virasoro blocks \cite{Harlow:2011ny, Fitzpatrick:2014vua, Hijano:2015rla, Banerjee:2016qca, Alkalaev:2015wia, Alkalaev:2018nik}. Let $\Psi_{W} (y, z_i)$ be an auxiliary $5$-pt conformal block with the degenerate operator $\psi(y)$  which corresponds to the vector $ |\psi \rangle$ \eqref{resc}.  Since in the classical limit $\Delta_{\psi}, Q_{\psi} \sim 1$ while other classical dimensions and charges are linear in $c$, this block is factorized into a product of the following form
\be
 \Psi_{W} (y, z_i) \Big|_{c \to \infty} \rightarrow \psi(y,z_i) \exp\left[\frac{c}{6} f_{W} (z_i)\right]\;.
\ee
The function $\psi(y,z_i)$ corresponds to the large-$c$ contribution of the operator  $\psi(y)$ into the classical conformal block. Due to the condition 
\eqref{3null} the conformal block $\Psi_{W} (y, z_i)$ satisfies the BPZ equation \cite{Belavin:1984vu}, which can be rewritten as a differential equation for the function $\psi(y,z_i)$
\be\label{BPZ}
\left[\frac{d^3}{dy^3} + 4 T(y,z_i) \frac{d}{dy} + \frac{2 d T(y,z_i)}{dy} - 4W(y, z_i) \right]\psi(y, z_i) =0\;.
\ee
The functions $T(y, z_i)$ and $W(y, z_i)$ read
\be
\label{cuur}
\ba{c}
\dps T(y, z_i) = \sum^{4}_{i=1} \left(\frac{\epsilon_i}{(y - z_i)^2}+\frac{c_i}{(y - z_i)}\right)\;, \\
\\
\dps W(y, z_i) = \sum^{4}_{i=1} \left(\frac{q_i}{(y - z_i)^3}+
\frac{a_i}{(y - z_i)^2}+\frac{b_i}{(y - z_i)}\right)~,
\ea
\ee
where we use notations \cite{Hulik:2018dpl}
\be 
a_i = \lim_{c \to \infty} \frac{6}{c} \frac{W^{(i)}_{-1} \Psi(y, z_i)}{\Psi(y, z_i)}\;, \qquad b_i = \lim_{c \to \infty}\frac{6}{c}\frac{W^{(i)}_{-2} \Psi(y, z_i)}{ \Psi(y, z_i)}\;, \qquad c_i = \partial_i f(z_i)\;. 
\ee
A few comments are in order. 
First, \eqref{BPZ} is the Fuchsian-type equation
with $4$ singular regular points. Second, assuming $q_i = 0$ and $a_i = b_i =0$ we see that the BPZ equation 
\eqref{BPZ} reduces to the equation  associated with the degenerate operator $V_{(3,1)}$ of the Virasoro algebra (see Appendix \bref{sec:V13} for details). Third, the conditions \eqref{asym}  allow fixing out 8 of 12 parameters $(a_i, b_i, c_i)$. Here we leave $(c_{_{2}}, a_1, a_2, b_2)$ independent. Simultaneously setting $(z_1, z_2, z_3, z_4) \rightarrow (1, z , 0, \infty)$ the expressions \eqref{cuur} become
\be
\label{pp}
T(y, z) = \frac{\epsilon_1}{(y-1)^2}  +\frac{\epsilon_2}{(y -z)^2} + \frac{\epsilon_3}{y^2} + \frac{c_2}{y-z} +  \frac{c_2 \left(z+ y-1\right)+\epsilon_2+\epsilon_3}{(1-y) y} \;,
\ee
\be
\label{ww}
\ba{c}
\dps W(y, z) = \frac{q_1}{(y-1)^3}+ \frac{q_2}{\left(y-z\right)^3}+\frac{q_3}{y^3} +  \frac{a_1}{(y-1)^2} + \frac{a_2}{\left(y-z\right)^2} + \frac{b_2}{y-z}
\\
\\
\dps-\frac{a_2 \left(2 z + y - 1\right)+a_1 (y+1)+b_2 \left(z \left(z + y - 1\right)+(y-1) y\right)+q_1+q_2+q_3+q_4}{(y-1) y^2}\;.
\ea
\ee
\paragraph{The monodromy problem for classical $\mathcal{W}_3$ blocks.} Let a string  $\; \psi_a(y, z), \; a = 1,2,3$ denote three independent solutions of \eqref{BPZ}. These solutions of the Fuchsian-type differential equation have non-trivial monodromy along contours enclosing singular points. Here we consider the contour $\Gamma$ bypasses the points $(1, z)$. The corresponding monodromy matrix $M_{ab}(\Gamma|z)$ is defined by
\be
\label{dm}
\psi_a (\Gamma \circ y, z) = \psi_b(y, z) M_{ab} (\Gamma|z)\;,
\ee
where $\Gamma \circ y$ denotes a traversal along the contour $\Gamma$. The matrix \eqref{dm} also depends on the functions $(c_{_{2}}, a_1, a_2, b_2)$ as parameters.

On the other hand, the main contribution to the monodromy matrix of the auxiliary $5$-pt block along the contour $\Gamma$ comes from the term $(z - y)^{\Delta_{\beta} - \Delta_{\alpha} - \Delta_{\psi}}$, where 
$\Delta_{\alpha}$ and $\Delta_{\beta}$ are intermediate dimensions of the $5$-pt block. Hence, the block must have
the monodromy matrix 
\be
\label{rest}
\tilde{M}_{ab} (\Gamma) = \exp[2 \pi i \gamma_a] \delta_{ab}\;, 
\ee
in a diagonal basis. The coefficients $\gamma_a$ are given by \eqref{gs}. In what follows, we always use the fusion rules \eqref{ngs}  for simplicity. 

The monodromy problem is to find solutions of the BPZ equation \eqref{BPZ} with the monodromy matrix \eqref{rest}. In other words, the eigenvalues of the monodromy matrices $M_{ab}$ \eqref{dm} and $\tilde{M}_{ab}$ \eqref{rest} must coincide. This imposes algebraic restrictions on the parameters $(c_{_{2}}, a_1, a_2, b_2)$, the so-called monodromy equations. Solving these equations and integrating the accessory parameter $c_{_{2}} = \partial_{z} f_{W}(z)$, we obtain the classical $\mathcal{W}_3$ block. It is important to mention that monodromy equations are generally not enough to determine all 4 parameters, which is consistent with the functional arbitrariness of the block. 

\subsection{Solution of the monodromy problem}
\label{sec:SMP}
\paragraph{HL approximation.} 

Since the solutions of the equation \eqref{BPZ} with arbitrary classical dimensions and charges are not generally known, we consider the heavy-light (HL) approximation.
More precisely, we assume that the classical dimensions and charges of the operators $\cO_{3}$ and $\cO_{4}$ are much larger than the remaining ones\footnote{ For the case of the Virasoro algebra, more than two heavy operators were also considered, see \cite{Alkalaev:2019zhs, Alkalaev:2020kxz}.
}
\be
\ba{c}
\epsilon_{1,2}, \tilde \epsilon  \ll \epsilon_H\;, \qquad q_{1,2} \ll q_{H}\;.
\ea
\ee
 By virtue of \eqref{2pt}, in the zeroth order of the HL approximation the classical dimensions and charges of heavy operators are related by $\epsilon_{3} = \epsilon_{4}$ and $q_3 = - q_{4}$, and henceforth will be denoted by $ \epsilon_{H} , q_{H}$ respectively. For the sake of simplicity, we also set $\epsilon_{1} = \epsilon_{2} \equiv \epsilon$ and $q_{1} = -q_{2} \equiv q$.

Now, we implement the heavy-light expansion 
\be
\label{decos}
\ba{c}
\psi(y, z) = \psi^{(0)}(y) + \psi^{(1)}(y,z)+...\;, \qquad
T(y, z) = T^{(0)}(y) + T^{(1)}(y,z)+...\;, \\
\\
W(y, z) = T^{(0)}(y) + W^{(1)}(y,z)+... \;,  \qquad f(z) = f^{(0)}(z) + f^{(1)}(z)+...\,,
 \ea
\ee
where zeroth-order parameters $a^{(0)}_{1,2}, b^{(0)}_{2}, c^{(0)}_{2}$ associated with heavy operators and the classical block $ f^{(0)}(z)$  are equal to 0. Because of that, we omit the superscript of the parameters, always referring to them in the first order of the HL approximation.

The monodromy matrix along contour $\Gamma$ is also expanded up to the first order in the HL approximation
\be
\label{dec}
M_{ab}(\Gamma|z) = M_{ab}^{(0)} (\Gamma) + M_{ab}^{(1)}(\Gamma|z) +...\;.
\ee

\paragraph{Monodromy matrix in the zeroth order.}
In zeroth order, the equation \eqref{BPZ} has the form
\be\label{BPZ0}
D^{(0)} \psi^{(0)}(y) =0\;, \qquad D^{(0)} = \left[\frac{d^3}{dy^3} + 4 T^{(0)} \frac{d}{dy} + \frac{2 d T^{(0)}}{dy} - 4W^{(0)}\right]\;, 
\ee
where 
\be
T^{(0)} = \frac{\epsilon_{H}}{y^2}\;,
\qquad W^{(0)} = \frac{q_{H}}{y^3}\;.
\ee
 There are three branches of solutions 
\be
\label{zs}
\psi_a^{(0)}(y) = y^{1+p_a}, \qquad a = 1, 2, 3\;,
\ee
where the exponents satisfy the characteristic equation of the third degree
\be
\label{che}
p^3_a - \alpha^2 p_a - 4 q_{H} = 0 \;, \qquad \alpha = \sqrt{1- 4 \epsilon_{H}} \;.
\ee
In order not to clutter up the calculations, we do not write explicit expressions for the roots of $p_a$. However, due to the Vieta's theorem for the cubic equation the exponents $p_i$ are satisfied by several relations, which are used below. For instance, $p_1 + p_2+p_3 = 0$ and $p_1 p_2 p_3 = 4 q_{H}$.  

The contour $\Gamma$ does not encircle the branch points of the solutions \eqref{zs} hence the monodromy matrix in the zeroth order is given by the identity matrix, $M_{ab}^{(0)} (\Gamma) = \delta_{ab}$. The monodromy matrix \eqref{rest}  in the zeroth order of the HL approximation coincides with $M_{ij}^{(0)} (\Gamma)$.  

\paragraph{First-order monodromy matrix  and monodromy equations.}
In the first order of the HL approximation, we have
\be
\label{BPZ1}
D^{(0)}  \psi^{(1)}(y) = - D^{(1)}  \psi^{(0)}(y)\;, \qquad D^{(1)} =  \left[4 T^{(1)} \frac{d}{dy} + \frac{2 d T^{(1)}}{dy} - 4W^{(1)}\right]\;,
\ee
where 
\be
\label{fot}
T^{(1)} =  \frac{c_2 (z-1)+2 \epsilon}{y-1}-\frac{c_2 z+2 \epsilon}{y}+\frac{c_2}{y-z}+\frac{\epsilon}{(z-y)^2}+\frac{\epsilon}{(y-1)^2}\;,
\ee
\be
\ba{c}
\label{fow}
\dps W^{(1)} = \frac{q}{(y-1)^3}- \frac{q}{\left(y-z\right){}^3} +  \frac{a_1}{(y-1)^2} + \frac{a_2}{\left(y-z\right){}^2} + \frac{b_2}{y-z}  \\
\\
\dps + \frac{\left(2 a_2 z - a_2 -a_1 + b_2 z (z -1)\right)(z-1)}{y^2} - \frac{2 a_2 \left(z-1\right)-2 a_1+b_2 \left(z-1\right){}^2}{y-1}\;.
\ea
\ee
 Note that equation \eqref{BPZ1} is the inhomogeneous differential equation with the same operator as \eqref{BPZ0} in the left hand side. Therefore, the first-order solutions  can be expressed in terms of zeroth-order solutions \eqref{zs} and take the form \cite{deBoer:2014sna}
\be
\label{s1}
\psi^{(1)}_{a} (y,z) = \psi_b^{(0)}(y) \int T_{ab}(y, z) dy\;,
\ee
where
\be
T_{ab} (y,z) = 2 \begin{pmatrix}
  \dps \frac{ 2 y \left( y W^{(1)} -  p_1  T^{(1)}\right)}{ p_{12} p_{13}} \;\;& \dps \frac{y^{1 - p_{12}} \left(p_3 T^{(1)} + 2 W^{(1)} y\right)}{p_{12} p_{13}} \;\;& \dps \frac{ y^{1 - p_{13}} \left(p_2 T^{(1)}+2 W^{(1)} y\right)}{p_{12} p_{23}}\\
  \\
\dps -\frac{ y^{1 + p_{12}} \left(p_3 T^{(1)}+2 W^{(1)} y\right)}{p_{12} p_{23}}\;\;&  \dps - \frac{ 2 y\left(  y W^{(1)} -  p_2 T^{(1)}\right)}{ p_{12} p_{13}} \;\;& \dps -\frac{ y^{1 - p_{23}} \left(p_1 T^{(1)}+2 W^{(1)} y\right)}{p_{12} p_{23}}\\
 \\
\dps \frac{ y^{1 + p_{13}} \left(p_2 T^{(1)}+2 W^{(1)} y\right)}{p_{13} p_{23}}\;\;& \dps \frac{ y^{1 + p_{23}} \left(p_1 T^{(1)}+2 W^{(1)} y\right)}{p_{13} p_{23}}\;\;& \dps \frac{ 2 y\left(  y W^{(1)} - p_3 T^{(1)}\right)}{ p_{13} p_{23}}
\end{pmatrix} \,, 
\ee and we introduce $p_{ij} \equiv p_i - p_j$. Hence, the monodromy matrix $M^{(1)}_{ab}$ has the form
\be
M^{(1)}_{ab} = \int_{\Gamma} dy T_{ab} \equiv I_{ab}\;.
\ee
Calculating the contour integrals using residues, one can obtain the elements of the monodromy matrix $M^{(1)}_{ab}$. In particular, the diagonal elements turn out to be 0, $M^{(1)}_{aa} = 0$. Other elements are linear functions of $a_{1,2}, b_2$ and $c_2$ and have the form
\be
\ba{c}
\dps \frac{I_{12}}{2 \pi i} = \dps \frac{ 2 z^{-p_{12}} \left( z \left(2 a_2 \left(2 - p_{12}\right)+c_2 p_3\right)+2 b_2 z^2-\left(p_{12}-1\right) \left( p_3 \epsilon +\left(p_{12}-2\right) q\right)\right)}{p_{12} p_{13}} \\
\\
\dps -\frac{2 \left( p_{12} \left(2 a_1+  p_3 \epsilon +\left(3-p_{12}\right) q\right)+4 a_2 z +2 b_2 z^2+c_2 p_3 z+  p_3 \epsilon  -2 q \right)}{p_{12} p_{13}}\;, \quad I_{13} = I_{12} \left( p_3 \leftrightarrow p_2\right)\;, \\
\\
\dps \frac{I_{21}}{2 \pi i} = - \frac{ 2 p_{12} \left(2 a_1+\epsilon p_3+\left(p_{12}+3\right) q\right)-4 a_2 z -2 b_2 z^2-c_2 p_3 x_2- \epsilon p_3+2 q}{p_{12} p_{23}} - \\
\dps \frac{2 \left(z^{p_{12}} \left( z\left(2 a_2 \left(p_{12}+2\right)+c_2 p_3\right)+2 b_2 z^2+\left(p_{12}+1\right) \left( \epsilon p_3 - \left(p_{12}+2\right) q\right)\right)\right)}{p_{12} p_{23}}\;, \quad I_{31} = I_{21} (p_2 \leftrightarrow p_3)\;, \\
\\
\dps \frac{I_{32}}{2 \pi i} = \frac{ 2\left( z \left(4 a_2+2 b_2 z +c_2 p_1\right)- p_{23} \left(2 a_1+\left(p_{23}+3\right) q\right)+ \epsilon p_1 \left(1 - p_{23}\right)-2 q \right) }{p_{23} p_{13}} \\
- \dps \frac{z \left(2 a_2 \left(p_{23}+2\right)+c_2 p_1\right)+2 b_2 z^2+\left(p_{23}+1\right) \left( \epsilon p_1 - \left(p_{23}+2\right) q\right)}{p_{23} p_{13}} \;, \quad I_{23} = I_{32} (p_1 \leftrightarrow p_3) \;.
\ea
\ee
Since the monodromy matrix of zero-order solutions is given by the identity matrix, in the first order the eigenvalues of the monodromy matrix $M^{(1)}_{ab}$  are given
\be
\label{om}
\tilde{M}^{(1)}_{ab} = \gamma^{(1)}_{a}  \delta_{a b}\;, \qquad \gamma^{(1)}_{1, 2} = \pm 4 \pi i \tilde \epsilon \;, \qquad \gamma^{(1)}_{3} = 0 \;.  
\ee
Eigenvalues of the  monodromy  matrix $M^{(1)}_{ab}$ should be equal to \eqref{om}.  It yields the monodromy equations of the following form
\be
\label{meq}
I_{12} I_{23} I_{31} + I_{13} I_{21} I_{32} = 0\;, \qquad I_{12} I_{21} + I_{13} I_{31} +  I_{23} I_{32} = - 16 \pi^2 \tilde \epsilon^2 \;.
\ee

A few comments are in order. First, the system \eqref{meq} consists of two equations for 4 parameters. Thus, the solution is parametrized by two functions which is a manifestation of the functional arbitrariness discussed in Section \bref{sec:cl}. Second, the monodromy matrix of the $4$-pt vacuum block ($\tilde \epsilon =0$) in the first order of the HL approximation is given by the zero matrix so we have 6 equations for 4 parameters. However, as shown in \cite{deBoer:2014sna}, not all of them are independent and in the case of the vacuum $4$-pt block the corresponding monodromy system has one solution. Integrating the accessory parameter $c_2$ one finds the block function \cite{deBoer:2014sna, Hegde:2015dqh}
\be
\ba{c}
\label{vb}
f^{(v)}_{W} (z) = \dps \frac{\left(3 q - \epsilon \right)}{2} \log \left(p_3 \left(2 z^{2 p_2+p_3}-z^{p_2+2 p_3}-1\right)+p_2 \left(z^{2 p_2+p_3}-2 z^{p_2+2 p_3}+1\right)\right) - \epsilon (1+p_1) \log z  \\
\\
- \dps \frac{\left(\epsilon+ 3 q\right)}{2} \left(\log \left(-p_3 \left(z^{p_2}-2 z^{p_3}+z^{2 \left(p_2+p_3\right)}\right)+p_2 \left(-2 z^{p_2}+z^{p_3}+z^{2 \left(p_2+p_3\right)}\right)\right)\right)\;.
\ea
\ee
The 4-pt vacuum Virasoro block is obtained by imposing the conditions $q=0=p_1$, $p_2 = - p_3 = \alpha$.  

\subsection{$4$-pt non-vacuum $\mathcal{W}_3$ blocks}

 There are several ways to exclude functional arbitrariness for the $4$-pt non-vacuum block. For example, one can fix the parameters $a_2 = 0, \; b_2 = 0$ and solve \eqref{meq} for $c_{2}$ and $a_1$.\footnote{More precisely, any pair of parameters can be fixed equal to 0, trivial configurations correspond to $c^{(n)}_2 = 0$.}  Since the integrals $I_{ab}$ are linear in $a_1$ and $c_2$, the monodromy system contains one third-order equation and one second-order equation. Isolating the accessory parameter $c_2$, we end up with a sixth order equation, solutions of which cannot be found in radicals. On the other hand, one can try to construct the solution as follows: for example, suppose that we fix 
\be
\label{ex}
I_{12} = I_{23}= I_{21} =0\;, 
\ee 
in the first equation of \eqref{meq} so the second one has the form 
\be
I_{13} I_{31} = - 16 \pi^2 \tilde \epsilon^2. 
\ee Solving \eqref{ex} one can find parameters $a_{1,2}, b_2$,  which are linear functions of $c_2$ and substitute to the equation above. It results in the quadratic equation in $c_2$ which can be solved but having a
too complicated form.

In this section, we consider another way of capturing functional arbitrariness. Since for the vacuum block all parameters $c_2, a_{1,2}, b_{2}$ are known, we consider the following ansatz for the $4$-pt non-vacuum block
 \be
 \label{mfab}
 c_2 = c^{(v)}_2 + c^{(n)}_2 \;, \qquad a_{1,2} =  a^{(v)}_{1,2}\;, \qquad  b_2 = b^{(v)}_2\;,
 \ee
where for the accessory parameter $c_2$ the superscript $v$ refers to a vacuum block and $n$ to a non-vacuum part of the block. Fixing the functional arbitrariness in this way, we assume that the intermediate operator with the classical dimension $\tilde \epsilon$ contributes only to the classical conformal block and does not change parameters $a_{1,2}, b_2$. 

Despite such a fixation of three parameters instead of two, one can see that under \eqref{mfab} the first equation in \eqref{meq} becomes trivial. The second equation is a quadratic equation in $c^{(n)}_2$ and its solution with the correct asymptotic takes the simple form 
\be
\label{an}
c^{(n)}_2 = \frac{\tilde \epsilon \dps}{2 z} \prod_{1\leq i \leq j \leq 3} p^{-1/2}_{ij} \left(\frac{ p^2_3\left(z^{ -p_{12}/2}-z^{ p_{12}/2}\right)^2}{p_{12}} - \frac{ p^2_2\left(z^{ - p_{13}/2}-z^{ p_{13}/2}\right)^2}{p_{13}} -  \frac{ p^2_1\left(z^{ - p_{23}/2}-z^{ p_{23}/2}\right)^2}{p_{23}}\right)^{-\frac{1}{2}},
\ee where $p_i$ are defined by \eqref{che}. 

Although we have not calculated this integral analytically, one can expand in $q_H/\alpha^2 \ll 1$. This approximation can be viewed as a deformation of the heavy-light Virasoro block \cite{Fitzpatrick:2014vua, Alkalaev:2015lca} by the classical spin-3 charge $q_H$. The zeroth term in the approximation corresponds to the non-vacuum part of the $4$-pt Virasoro heavy-light block. The exponents $p_i$ \eqref{che} take the form
\be
p_1 = \frac{4 q_H}{\alpha^2} + ...\;, \qquad p_{2,3} = \pm \alpha + \frac{2 q_H}{\alpha^2} + ... \;,
\ee
and $f^{(n)}_{W}$ reads
\be
\label{finr}
f^{(n)}_{W} = \tilde \epsilon \left( f_0 + \frac{q^2_{H}}{4\alpha^5} f_{2} + ...\right)\;,  
\ee
where 
\be
\label{nc}
\ba{c}
f_0 = 2\text{Arcth} \left(z^{\alpha/2}\right)\;, \\
\\
f_2 = 27  \left(\text{Li}_2\left(z^\alpha\right)-4 \text{Li}_2\left(z^{\alpha/2}\right)\right)  \log z +27 \alpha \; \text{Arcth} \left(z^{\alpha/2}\right)  \log ^2 z +\dps \frac{8 (z^{\alpha/2}+z^{-\alpha/2})}{\alpha} + \\
\\
\dps 27 \alpha^{-1} z^{\alpha/2} \Phi \left(z^\alpha,3,\half\right) + \frac{9\alpha  (z^{3 \alpha /2 } +z^{\alpha/2 } ) \log^2 z}{(1-z^\alpha)^2}  + \frac{84 z^\alpha \log z}{1-z^\alpha}\;. 
\ea
\ee
Here $\Phi(x ,s,a)$ denotes the Lerch transcendent function (see \cite{Guillera_2008} for review)
\be
\Phi (x,s,a)=\sum_{k=0}^{\infty } \frac{x^k}{(a+k)^s}\;,
\ee
which generalise the polylogarithm given by $ \dps \text{Li}_s (x) = \sum_{k=1}^{\infty} \frac{x^k}{k^s}$. Next orders in $q_H/\alpha^2$ can be found in exactly the same way. 
Thus, the result for the non-vacuum 4-pt $\mathcal{W}_3$ block can be represented as
\be
\label{afr}
\ba{c}
f_{W} = -\epsilon (1-\alpha) \log z - 2 \epsilon \log \left(z^{-\alpha/2}-z^{\alpha/2}\right) + \tilde \epsilon f_0 + \\
\\
\dps \frac{ q_{H} \left(-4 \alpha \epsilon \left(z^{\alpha}-1\right)^2 \log z -18 q \left(z^{2 \alpha}-1\right)+6 q \alpha  \left(\left(z^{\alpha}+4\right) z^{\alpha}+1\right) \log z \right)}{\alpha^3 \left(z^{\alpha}-1\right)^2} + \frac{q^2_{H}}{4\alpha^5} f_{2} +....
\ea
\ee
Few comments are in order. First, we have highlighted the $4$-pt non-vacuum Virasoro block in the first line, and the corrections to it in the second line, where $f_0$ and  $f_2$ are determined by \eqref{nc}. For $\tilde \epsilon =0$ we get the vacuum block \eqref{vb} in the approximation $q_H/\alpha^2 \ll 1$. Second, note that in \eqref{afr} the first-order corrections in $q_H/\alpha^2$ are determined by the vacuum block \eqref{vb}, which does not contribute to the second order. Third, the block \eqref{afr} can be rewritten as a function $z^{\alpha/2}$,
which indicates the presence of holographic variables \cite{Fitzpatrick:2015zha, Alkalaev:2020kxz}. 

The study of the non-vacuum block was partly motivated by the connection between the $\mathcal{W}_N$ vacuum block and the  Virasoro blocks (both vacuum and non-vacuum) \cite{Hegde:2015dqh}. More precisely, the basis  states formed the $\mathcal{W}_N$ module can be rewritten in terms of Virasoro states. However, based on \eqref{afr}, it is still not entirely clear whether the $\mathcal{W}_3$ non-vacuum block can be organized in the same way.

\section{Conclusion}
\label{sec:conclusion}

In this work, the monodromy method for the $4$-pt non-vacuum classical block associated with the $\mathcal{W}_3$ algebra was considered. Using the fusion rules for the auxiliary $5$-pt block, we have obtained monodromy equations that partially define the $4$-pt classical block. Having fixed the functional arbitrariness in the way mentioned above, we have found the $4$-pt non-vacuum block. We also studied the reduction of the monodromy method for the $\mathcal{W}_3$ algebra to one for the Virasoro algebra and showed that the results obtained within its framework go over to the well-known expressions for the $4$-pt non-vacuum Virasoro block. 

It would be of interest to apply the monodromy method to conformal blocks, where the functional arbitrariness is fixed by the choice of special primary operators \cite{Fateev:2007ab, Belavin:2016qaa} or within the framework of the isomonodromic approach \cite{Gavrylenko:2015wla}. Another interesting topic is to consider blocks where one or two primary operators are replaced by  degenerate operators with a highest-weight in the fundamental representation of $sl(3)$. 

In this work, we did not touch upon the holographic interpretation of the results obtained. It would be interesting to study dual objects in terms of Wilson lines and establish a correspondence similar to the examples given in \cite{Besken:2016ooo}. 

\vspace{5mm}

\noindent \textbf{Acknowledgements.} I thank Konstantin Alkalaev, Vyacheslav Didenko, Alexey Litvinov and Sylvain Ribault for useful discussions. The work was supported by the Russian Science
Foundation grant 18-72-10123.

\appendix

\section{Monodromy method for the $4$-pt Virasoro block via $V_{(3,1)}$}
\label{sec:V13}
Here we elaborate the monodromy method associated with the degenerate operator $V_{(3,1)}$ of the Virasoro algebra.\footnote{We thank M.A. Vasiliev for pointing out this aspect.} 

\paragraph{Fusion rules.} The null-vector associated with the degenerate operator $V_{(3,1)}$ with $\Delta_{(3,1)} \equiv \Delta$ is given by
\be
\label{sv}
|\psi_{3v} \rangle = \left(L_{-3} - \frac{2}{\Delta + 2} L_{-1} L_{-2} + \frac{1}{(\Delta+1)(\Delta+2)} L^3_{-1} \right)| \Delta \rangle =0\;, 
\ee
where $\Delta = -1 - \dps \frac{12}{c}$ at $c \rightarrow \infty$. Considering the $3$-pt function $V_{\alpha \beta \psi_{_{3v}}} =0$ with the operator $\psi_{_{3v}}$ corresponding to \eqref{sv}, we obtain the following fusion rule
\be
(\Delta_{\alpha} - \Delta_{\beta}) (\Delta^2_{\alpha} - (\Delta_{\beta} - \Delta)(1 - \Delta) - \Delta_{\alpha}(1 + \Delta_{\beta} + \Delta)) = 0\;.
\ee
Solving this equation with respect to the variable $\gamma = \Delta_{\beta} - \Delta_{\alpha} - \Delta$ in the classical limit, we find
\be
\label{vf}
\gamma_1 = 1 \;, \qquad \gamma_{2,3} = \dps  \left(1 \pm  \sqrt{1 - 4 \epsilon_{\alpha}} \right)\;. 
\ee

\paragraph{BPZ equation.} 
The BPZ equation associated with the singular vector \eqref{sv} has the following form
\be
\label{BPZV}
\left[\frac{d^3}{dy^3} + 4 T(y,z_i) \frac{d}{dy} + \frac{2 d T(y,z_i)}{dy} \right]\psi(y, z_i) =0\;,
\ee
where $ T(y,z_i)$ is defined in the first line of \eqref{cuur}. The equation can be considered as a special case of \eqref{BPZ} at $W(y,z_i)=0$. Using the HL approximation described in Section \bref{sec:SMP}, we see that the zeroth-order equation takes the form
\be
\label{ve13}
\left[\frac{d^3}{dy^3} + \frac{4 \epsilon_H}{y^2} \frac{d}{dy} - \frac{4 \epsilon_H}{y^3} \right]\psi^{(0)}(y) = 0\;,
\ee
and solutions are
\be
\label{ZV}
\psi_{1,2}^{(0)}(y) = y^{1 \pm \alpha}, \qquad \psi_{3}^{(0)}(y) = y \;.
\ee
Note that this equation is closely related to the previously studied BPZ equation, which is associated with the degenerate operator $V_{(2,1)}$ \cite{Fitzpatrick:2014vua, Alkalaev:2018nik}. In the zeroth order, the equation and its solutions read
\be
\label{v21}
\left[\frac{d^2}{dy^2} +  \frac{ \epsilon_H}{y^2} \right] \phi^{(0)}(y) = 0\;, \qquad \phi^{(0)}(y) = y^{\frac{1\pm \alpha}{2}} \;. 
\ee
We see that the solutions \eqref{ZV} can be constructed as quadratic combinations of solutions to the equation \eqref{v21}. 

Due to the role of zero-order solutions, the monodromy properties of the equation \eqref{ve13} are governed by ones of the equation \eqref{v21}. More precisely, the monodromy matrix of the first-order solutions of \eqref{BPZV} has the form
\be
I_{ab} = \frac{2 \pi i}{\alpha} \begin{pmatrix}
  \dps 0 \;\;& \dps 2 I_1  \;\;& \dps -2 I_2 \\
\dps - I_2 \;\;&  \dps 0 \;\;& \dps 1\\
\dps I_1\;\;& \dps 0 \;\;& 0 
\end{pmatrix} \,, \qquad \alpha = \sqrt{1 - 4 \epsilon_H}\;, 
\ee
where
\be
I_1 = c_2 z \left(z^{\alpha}-1\right)+\epsilon \left((\alpha+1) z^{\alpha}+\alpha-1\right) \;,\qquad
I_2 = I_1 (\alpha \rightarrow - \alpha) \;.
\ee
Comparing the eigenvalues of this matrix and \eqref{vf} with $\epsilon_{\alpha} = \tilde \epsilon$ at $\tilde \epsilon \ll \epsilon_{H}$, we obtain one non-trivial monodromy equation
\be
I_1 I_2 = - \alpha^2 \tilde \epsilon^2\;, 
\ee
which can be solved with respect to $c_2$. After integrating this accessory parameter, we get
\be
f(z) =  2 \tilde \epsilon  \text{Arcth} \left(z^{\alpha/2}\right)-2 \epsilon \log \left(1-z^{\alpha}\right)+ \epsilon (\alpha-1)  \log z\;, 
\ee
what constitutes the $4$-pt non-vacuum Virasoro block. Due to the fact that the result for the classical block should not depend on the choice of the degenerate operator from which the auxiliary block is constructed, similar reasoning is applicable for any degenerate operator of the Virasoro algebra $V_{(r,s)}$, see \cite{Hou:2022hxv} for recent studies. 
\providecommand{\href}[2]{#2}\begingroup\raggedright\endgroup

\end{document}